# Basis function compression for field probe monitoring

Paul I. Dubovan[1,2], Gabriel Varela-Mattatall[1,3], Eric S. Michael[4], Franciszek Hennel[4], Ravi S. Menon[1,2], Klaas P. Pruessmann[4], Adam B. Kerr[5,6], Corey A. Baron[1,2]

[1]Department of Medical Biophysics, Western University, London, Ontario, Canada

[2]Centre for Functional and Metabolic Mapping, Western University, London, Ontario, Canada

[3]Lawson Health Research Institute, London, Ontario, Canada

[4]Institute for Biomedical Engineering, ETH Zurich and University of Zurich, Zurich, Switzerland

[5]Center for Cognitive and Neurobiological Imaging, Stanford University, Stanford, California, USA,

[6]Department of Electrical Engineering, Stanford University, Stanford, California, USA

**Conflict of Interest**

Klaas Paul Pruessmann holds a research agreement with and receives research support from Philips and is a shareholder of Gyrotools LLC.

* Corresponding Author
Name: Paul I. Dubovan
Address: Centre for Functional and Metabolic Mapping
Western University
1151 Richmond St.
London, Ontario, Canada
N6A 3K7
Tel: 519-992-7794
E-mail: pdubovan@uwo.ca

**Funding Information**

NSERC Discovery Grant, Grant Number: RGPIN-2018-05448; Canada Research Chairs, Number: 950-231993; Canada First Research Excellence Fund to BrainsCAN; NSERC CGS D program

**Word Count**

4974

Submitted to Magnetic Resonance in Medicine

**Abstract**

**Purpose:** Field monitoring using field probes allows for accurate measurement of magnetic field perturbations, such as from eddy currents, during MRI scanning. However, errors may result when the spatial variation of the fields is not well-described by the conventionally used spherical harmonics model that has the maximum order constrained by the number of probes. The objective of this work was to develop and validate a field monitoring approach that compresses higher order spherical harmonic basis functions into a smaller set of new basis functions that can be computed from fewer probes.

**Methods:** Field monitoring of acquisitions was repeated with probes in different locations. High-order field dynamics were computed from this "calibration" probe data assembled from all scans, from which compression matrices could be devised using principal component analysis. Compression matrices were then utilized to fit field dynamics using "compressed" basis functions with data from 16 probes, which were then used in image reconstruction. Performance was evaluated by assessing the accuracy of computed field dynamics as well as *in vivo* image quality. Technique generalizability was also assessed by using various acquisition and diffusion encoding strategies for the calibration data.

**Results:** Qualitative and quantitative improvements in accuracy were observed when using the proposed fitting method in comparison to the conventional approach. However, compression effectiveness was influenced by the specific acquisition data included in the calibration set.

**Conclusion:** The ability to tailor basis functions to more compactly describe the spatial variation of field perturbations enables improved characterization of fields with rapid spatial variations.

# 1 | Introduction

Field monitoring is an effective tool for measuring field perturbations during acquisitions due to its high sensitivity to spatial and temporal field variations.[1–4] This technique enables the accurate measurement of spatially invariant field offsets, first-order gradients, and higher order field deviations arising primarily from gradient-induced eddy currents.[5,6] Of note are eddy currents generated by the strong diffusion gradients used in diffusion encoding sequences. These include pulsed gradient spin echo (PGSE),[7] oscillating gradient spin echo (OGSE),[8–10] and spherical tensor encoding schemes,[11,12] which are notorious for introducing large field deviations.[13] When unaccounted for, these fields can produce notable image artifacts and errors in computed diffusion metrics such as mean diffusivity and fractional anisotropy (FA),[14] as well as in advanced parameters like kurtosis[15] and microscopic fractional anisotropy ($\mu$FA).[16–18] To account for these effects, field monitoring measurements can be incorporated in image reconstruction strategies, thereby reducing artifacts and improving the accuracy of associated diffusion metrics.[19,20]

The most widely used commercial field monitoring system (Skope, Zurich) uses 16 field probes that record the local MR signal phase, which allows computation of the local magnetic field. This allows for characterization of spatial variations using real-valued spherical harmonics up to 16 basis functions, corresponding to third order.[1] For most field monitoring uses, this is not problematic as fitting to second or third order is sufficient to capture most field perturbations. However, in some cases phase contributions originating from higher orders may be non-negligible, such as when using head-only MRI scanners or high-performance gradient systems that operate at very high gradient strengths and slew rates.[21–23] In addition, scenarios may exist where field probes are located in regions that experience rapidly-varying eddy current modes and gradient nonlinearity, which may not be well characterized by spherical harmonics fit using only 16 probes.[24,25] This can lead to biased or erroneous fitting of the phase amongst the available spatial orders, and in turn corrupted image quality. To overcome these challenges, fitting the basis functions using more field probes than functions leads to an overdetermined system, and hence a better conditioned least-squares problem that reduces the error in calculated coefficients.[1] Integrating more probes also permits the characterization of higher spatial orders, which may help to more accurately fit the probe phase if higher orders are present. While the number of probes is limited by hardware and available space, the generation of additional field probe measurements can be accomplished by repeating the acquisition after moving the probes into different positions, from which the probe data from the separate scans can be compiled together to form a larger probe array. Although advantageous, this approach is time consuming and not possible for concurrent field monitoring given the subject's presence in the coil.

In this work, we propose and test a "basis function compression" strategy to accurately calculate high-order field dynamics from a limited number of field probes. Using a head-only 7T MRI, we demonstrate greatly improved reconstruction quality for a transmit-receive coil with 16 integrated field probes where all probes extend into the nonlinear region of the gradient.

## 2 | Methods

The compression technique has been made publicly available (see Data Availability Statement).

### 2.1 | Determination of compressed k-coefficients

#### 2.1.1 | Establishing ground truth calibration k-coefficients

In field monitoring, the phase can be described as separable in space and time and can be expressed as the product of the probe-position-sampled spherical harmonics and the "k-coefficients" that describe the time-dependence of each basis function. For more details on the derivation of phase coefficients using field probes, please refer to the author's previous work,[24] and the sections described in Barmet et al.[1] and Wilm et al.[19] Here, we consider a "calibration scan" that consists of multiple acquisitions with various gradients applied to excite a variety of eddy current modes, such that a total of $N_t$ samples are acquired across all time points and gradient waveforms. To increase the total number of field probe positions, this scan is repeated multiple times with probes in different locations, arriving at a total of $N_p$ probe locations. The phase accrual of the probes is given by:

$$\phi_{calib} = P_{calib} k_{calib} \quad (1)$$

where $\phi_{calib}^{N_p \times N_t}$ is the phase, $P_{calib}^{N_p \times N_b}$ is the "probing matrix" that contains $N_b$ solid harmonic basis function values at the location of each probe, and $k_{calib}^{N_b \times N_t}$ are the k-coefficients that correspond to each basis function. Accordingly, $k_{calib}$ can be determined from the calibration data $\phi_{calib}$ using a Moore-Penrose pseudoinverse of $P_{calib}$.

#### 2.1.2 | Basis Function Compression

The goal of basis function compression is to more compactly represent the typical field perturbations experienced on a system compared to the solid harmonic basis functions that are typically used. Similar to Wilm et al.,[21] a set of compressed k-coefficients can be determined as weighted combinations of solid harmonic k-coefficients using an economic singular value decomposition (SVD) of the ground truth harmonic coefficients determined from the calibration scan. Since typical gradient systems are designed to generate linearly varying phase distributions, it is reasonable to assume *a priori* that the 0[th] and 1[st] order solid harmonics are effective basis functions that can thus be omitted from the SVD:

$$\Gamma^{high} k_{calib}^{high} = U \Sigma V^T \qquad (2)$$

where the "high" superscript denotes that only 2nd and higher order terms are included. $\Gamma^{high}$ is a diagonal matrix of weights that reduces bias in the SVD from signal and noise amplification that can occur from the different scalings of each order. In this work we set $\Gamma_{i,i}^{high} = (0.1)^{l_i}$, where $l_i$ is the order of the $i$'th basis function, having $i = 1$ start at the 2nd order basis functions. This choice of $\Gamma$ is equivalent to weighting the basis functions the same over a standard imaging volume with a 10-cm radius, or to converting the units of k-coefficients from rad/m$^l$ to rad/dm$^l$.

The singular vectors in $U$ describe linear combinations of k-coefficients that most compactly describe the calibration data. Thus, a compression matrix can be defined by omitting the columns corresponding to the lowest singular values, similar to algorithms for coil compression:[26,27]

$$C^{high} \equiv U_{i,j}, \ 0 < i < (N_b - 4), 0 < j < L \quad (3)$$

where the "high" subscript denotes this compression matrix corresponds to the harmonic orders of 2 and higher, $L$ is the number of singular vectors retained (which may be based on singular value thresholding), and 4 is subtracted from $N_b$ because the 0th and 1st order terms were omitted in Equation (2). To create weighting and compression matrices that can be applied to all orders, block diagonal matrices can be constructed as follows:

$$\Gamma = \begin{bmatrix} I & 0 \\ 0 & \Gamma^{high} \end{bmatrix} \text{ and } C = \begin{bmatrix} I & 0 \\ 0 & C^{high} \end{bmatrix} \qquad (4)$$

where I is a $4 \times 4$ identity matrix that retains the 0th and 1st order k-coefficients. Accordingly, k-coefficients for an arbitrary scan can be compressed via:

$$\hat{k} = C^T \Gamma k \quad (5)$$

Likewise compressed k-coefficients can be uncompressed via:

$$k \approx \Gamma^{-1} C \hat{k} \quad (6)$$

where the equality is approximate due to the discarding of some singular vectors. Compared to the full set of k-coefficients, there are a smaller number of compressed k-coefficients and thus fewer probes would be needed to compute them. For routine image reconstruction using field monitoring measurements from a conventional 16-probe array, a maximum of 12 higher order compressed basis functions is permitted, given that 4 basis functions are allocated to the uncompressed 0th and 1st order terms. Substituting Equation (6) into (1) to recast Equation (1) into a form with compressed k-coefficients yields:

$$\phi = P\Gamma^{-1}C\hat{k} \quad (7)$$

where $P$ and $\hat{k}$ represent the probing matrix and compressed k-coefficients, respectively. Equation (7) can be observed to have a similar format as Equation (1) with the substitution:

$$\hat{P} \equiv P\Gamma^{-1}C \quad (8)$$

which leads to $\phi = \hat{P}\,\hat{k}$. Here, $\hat{P}$ can be interpreted as representing a compressed set of basis functions that correspond to the compressed k-coefficients. Accordingly, the compressed k-coefficients can be determined from the phase data using the pseudoinverse of $\hat{P}$:

$$\hat{k} = (\hat{P}^T\hat{P})^{-1}\hat{P}^T\phi \quad (9)$$

After computing $\hat{k}$ from Equation (9), Equation (6) can be used to retrieve the k-coefficients in their original uncompressed solid harmonic form, which can then be used in existing expanded encoding reconstruction pipelines that utilize these basis functions.

A detailed diagram of the process is illustrated in Figure 1.

## 2.2 | Scan Details

Scanning was conducted on a 7T head-only MRI scanner (Siemens MAGNETOM Terra Plus) equipped with an AC-84II head gradient coil (80 mT/m max gradient and 400 T/m/s max slew rate) at Western University's Centre for Functional and Metabolic Mapping. This study was approved by the institutional review board, and informed consent was obtained before scanning. Field monitoring was performed using a commercial Clip-On Camera (Skope, Zurich, Switzerland) consisting of 16 probes integrated into a 32-channel receive, 8-channel transmit RF head coil.[25]

### 2.2.1 | Calibration Acquisition Details

Field monitoring measurements from different probe locations were performed by rotating the field probe coil about the z-axis and translating the coil along the z-direction, where identical acquisitions were performed in each location with no imaging subject. 9 different coil orientations were used: 3 z-positions (spaced approximately 4-5 cm from each other), and three equiangular rotated positions for each given z-location. From this collection of probe positions, a synthetic probe array consisting of 100 probes was generated using probes positioned up to a maximum Euclidean distance of 16 cm from isocenter, as well as within individual x, y, z distances of 14 cm from isocenter to avoid excessive field inhomogeneity and gradient nonlinearity. For this experiment, $5^{th}$ order fits were performed using the calibration data with Equation 1 to establish ground truth k-coefficients. Unless stated otherwise, $k_{calib}$ included data from all acquisition time points, slices, $b_0$ and diffusion weighting volumes.

The success of compression depends on the ability of the calibration data to accurately represent all the field perturbations that might be encountered in scans that the compression matrix is applied to. Assuming that the dominant source of perturbations is gradient-induced eddy currents, the calibration data should include various gradient waveforms. On one extreme, calibration data could contain nearly all possible gradient waveforms, such as a collection of chirped or triangular pulses typically used to calibrate gradient impulse response functions.[28,29] Alternatively, one could use calibration data that is very similar to the desired acquisitions, such as a collection of spiral acquisitions at various resolutions to calibrate for spiral acquisitions only. The former approach has the advantage of generalizability, but it may not be very compressible compared to the latter. To explore these trade-offs, various calibration scans were performed (Table 1). Notably, any number of these calibration scans can be combined to obtain a single compression matrix from Equations 1-4, where combining different scans is expected to improve generalizability at the expense of compressibility. All diffusion tensor imaging (DTI) acquisitions consisted of 2 b = 0 s/mm² acquisitions and 30 diffusion directions, which were uniformly distributed using electrostatic repulsion of particles on a sphere.[30] Other common imaging parameters include FOV: 192 x 192 mm², slice thickness: 2 mm, number of slices: 10, axial orientation, rate 2 undersampling, scan time approximately 1.5 minutes for each coil location. Differing scan parameters are described in Table 1.

### 2.2.2 | *In vivo* Imaging

One healthy volunteer was scanned with scans 1 to 4, and field monitoring was performed concurrently.[32] In this baseline position, the mean distance between field probes and isocenter is 13.5 cm (range: 11.9–14.9 cm), with all probes being 8–35% outside the 22 cm DSV. A second healthy volunteer was scanned using a single-shot spiral b-tensor encoding acquisition[11,12] with concurrent field monitoring. The following imaging parameters were used: FOV: 200 x 200 mm², 1.5 mm isotropic resolution, 88 slices, TE/TR: 82/10,500 ms, undersampling rate 3, bandwidth = 2194 Hz/Px. The protocol consisted of a 94-direction diffusion scheme with 6 b = 0 s/mm² volumes, 6, 26, and 26 linear tensor-encoded (LTE) volumes at b = 150, 1000, and 2000s/mm², respectively, and 30 spherical tensor-encoded (STE) volumes at b = 2000s/mm², totaling a scan time of 16.5 minutes.

For all *in vivo* scans, cartesian dual-echo gradient-echo acquisitions were used to estimate $B_0$ maps for inclusion in a model-based reconstruction to correct for static off-resonance effects. The imaging parameters were as follows: FOV = 210x210 mm$^2$, 2 mm isotropic resolution, 74 slices, TE$_1$/TE$_2$ = 4.08/5.10 ms, TR = 542 ms.

## 2.3 | Image Reconstruction

Image reconstruction was performed using an iterative expanded encoding model-based reconstruction[19] in MATLAB via the MatMRI toolbox.[33] All images were reconstructed using the conjugate gradient method with Tikhonov regularization and a regularization weighting of 0.2. All reconstructions converged within 10-30 iterations. Coil compression to 24 virtual coils was performed to improve reconstruction speed.[26,27,34,35] ESPIRiT was used to estimate sensitivity coil maps.[36] Synchronization delay between the MRI and field probe data was corrected using an automatic, retrospective algorithm.[37] Noise correlation between receivers was corrected using prewhitening before any reconstructions.[38] Field dynamics were also adjusted to account for vendor Maxwell corrections and eddy current compensation that are measured by the field probes, using methods described previously by the authors.[25] Unless stated that "uncorrected" k-coefficients were used, all k-coefficient fitting performed for data acquired on our scanner, even when implementing conventional fits, included a weighting function $W$ that suppresses errors from distal field probes, as described in previous work.[24] Accordingly, $P$ is replaced by $WP$ in the equations, producing $\phi = WPk$ and $k = (P^T W^2 P)^{-1} P^T W^2 \phi$ for k-coefficient calculation (likewise for $\hat{P}$). Similar to previous work,[24] concomitant fields were included in all k-coefficient fitting, including compressed fitting, by iterative computation of concomitant gradient phase from the linear k-coefficients after evaluating Equation 9, removal of concomitant phase from $\phi$, followed by re-evaluation of Equation 9.

Following image reconstruction, FA maps of *in vivo* images were computed using the MRtrix3 package.[39] Additionally, b-tensor encoding volumes were denoised by performing PCA denoising on the complex data,[40] and ADC and μFA maps were calculated using matMRI.[41]

## 2.4 | Data analysis

To quantitatively assess basis function compression performance, k-coefficient root-mean-square-error (RMSE) was calculated from the concurrently monitored acquisitions with 16 probes with respect to the ground truth k-coefficients, $k_{GT}$, computed from the full 100 probe locations from the calibration scans. Earlier work has shown that inadequate modeling of the spatial variation of field dynamics causes the omitted higher orders to be projected to the lower orders, and that it is errors in the lower order k-coefficients that are primarily responsible for image artifacts.[24] Accordingly, RMSE was only computed for the first 9 basis functions, corresponding to the 0[th] to 2[nd] order. For similar rationale as in section 2.1.2, $\Gamma_{i,i} = (0.1)^{l_i}$ was applied as well, except here the basis functions start at the 0[th] order:

$$RMSE = \sum_{i=1}^{9} \sqrt{\frac{1}{N_t} \sum_{j=1}^{N_t} \left( \Gamma(k_{i,j} - k_{i,j}^{GT}) \right)^2} \qquad (10)$$

The RMSE was computed separately for each volume, followed by calculation of the mean and standard deviation across all volumes. Qualitative assessments were performed from image reconstructions and, similar to Wilm et al.,[21] for phase maps from compressed basis functions in planes orthogonal to two cardinal axes.

## 2.5 | Technique Performance: Additional sites

To assess the performance of the technique on different infrastructure, field monitoring data was collected from ETH Zurich (3T Philips Achieva scanner, 200 mT/m max gradient and 600 T/m/s max slew rate, ~20-cm imaging volume[42]) and Stanford (GE 3T UHP scanner, 100 mT/m max gradient, 200 T/m/s max slew rate, 50-cm imaging volume). The ETH Zurich acquisition implemented single-shot spiral diffusion-weighted imaging and consisted of both *in vivo* and field monitoring measurements, like the Western University data. Conversely, Stanford data consisted of an acquisition of chirped gradient pulses, and only considered field monitoring measurements. Specific details are described in Supporting Information S1.

## 3 | Results

Performing reconstructions informed by third and fifth order fits calculated using the 100-probe array showed substantial improvements in DWI and FA maps over reconstructions informed by third order fits calculated conventionally using the 16-probe arrangement. Incremental improvements were observed when comparing 5[th] to 3[rd] order fit with 100 probes (Video S1).

When truncating the compression matrix to different numbers of singular vectors when using scan 1 for both calibration (100 probes) and k-coefficient calculation from probe measurements during the *in vivo* scan (16 probes), the RMSE plot and resulting images showed the lowest RMSE and best image quality when using 5 singular values (Figure 2a,b & Figure S1). Using 5 singular values, computed compressed fifth order-fit k-coefficient plots exhibited the lowest RMSE (Figure 2c), showing better similarity for first order terms when compared with a conventional first order fit (Figure S2). Mean DWI and FA maps reconstructed from a compressed fifth order fit were of comparable quality to images reconstructed from ground truth k-coefficients computed from the scan 1 calibration data, while image quality was significantly degraded when using conventional fitting schemes, especially when increasing the fitting order (Figure 2d). The total number of terms used in the compressed fit (uncompressed lower order and compressed higher order) equaled 9, which is equivalent to the total number of terms included in a $2^{nd}$ order fit, yet the compressed $5^{th}$ order fit performed markedly better.

Different gradient profiles were observed for spiral trajectories acquired with different resolutions (Figure 3a), and frequency content analysis showed significant separation of power spectral densities across the x and y gradient channels (Figure 3b), which could affect the spatial dependence of eddy current modes excited in nearby hardware. The combination of both resolutions into the compression matrix resulted in maps most comparable to the higher resolution case (Figure 3c). Both resolutions had the lowest RMSE when the same scan was used to generate calibration data, but the combined calibration data performed well for both resolutions. The RMSE was not very sensitive to choice of calibration data for the 2-mm resolution scan (Figure 3d), likely due to the gradient spectra of the 1.3-mm resolution scan having high overlap with the whole 2-mm resolution scan. Little qualitative differences in DWI were seen at either resolution, regardless of the calibration data used for compression.

Figure 4a shows that strong similarity in basis function maps was maintained as the number of diffusion directions was reduced from 30 to 6, but a complete mismatch was observed when only one diffusion direction was included in the calibration. RMSE analysis was performed for diffusion-calibration data that incorporated 1 to 30 diffusion directions and was assessed for $b_0$-only acquisitions (Figure 4b), as well as for all acquisitions (Figure 4c). For this specific analysis, calibration data did not include $b_0$ acquisitions given that the assessment of diffusion direction quantity was of interest. In both cases, RMSE increase was only observed below 6 directions. Similarly, image quality was similar for calibrations using 30 and 6 directions, which drastically lowered when using a single direction (Figure 4d).

Very little similarity in compressed basis function maps was observed between PGSE and OGSE calibration data (Figure 5a), yet there was strong similarity between the combined and PGSE cases. Accordingly, RMSE was highest for PGSE acquisitions that used OGSE calibration data, while the combined calibration data performed relatively well for both cases. (Figure 5b). These errors were observed as subtle blurring in the OGSE-calibrated mean PGSE DWI (Video S2; Figure 5c).

Different gradient profiles were observed for the spiral and EPI trajectories (Figure 6a), and frequency content analysis showed relatively larger spectral energy only along the x-channel for the EPI trajectory, overlapping significantly with both channels of the spiral trajectory (Figure 6b). There was strong similarity between the combined case and the EPI trajectory (Figure 6c). RMSE values were the lowest when using calibration data matching the image reconstruction (Figure 6d). The combined calibration performed well for EPI, but poorly for spiral. Small but noticeable qualitative differences in mean DWI were observed when calibrating with the other trajectory, or for calibrating with the combined data for spiral (Figure 6e).

The b-tensor encoding scan reconstructed using the calibration matrix determined from scan 1 resulted in DWIs and ADC/µFA maps with high quality throughout the brain (Figure 7a). The high-quality STE images suggest good generalizability from LTE encodings used in the calibration. Mean DWI acquired from the LTE encoding scheme exhibited large improvements in image quality in comparison to images that were reconstructed with the conventional second order fit (Figure 7b).

Analysis of k-coefficient profiles calculated from the ETH Zurich scan showed good agreement between compressed fifth order terms and ground truth fifth order terms (Figure S3a), and a significant reduction in overall error was observed when using the proposed fitting method as opposed to conventional third order fit (Figure 8a). Accordingly, mean DWI and FA maps informed by compressed fifth-order k-coefficients showed better agreement with ground truth reconstructions, than conventional third order fits (Figure 8b). For the frequency-sweep acquisition from Stanford, compressed fifth order k-coefficient data also showed better agreement with the ground truth profiles when compared to conventional third order (Figure S3b), which resulted in an overall reduction in mean RMSE error (Figure 8d). That said, the reduction in RMSE is modest, likely because there are few higher order terms for the large body gradient compared to the other specialized gradient systems investigated in this work. A combination of all the frequency-sweeps from all three gradient channels provided the best compression performance, as opposed to calibrations of individual channels (Figure 8e).

Compression performance based on the probe quantity showed incremental increases in RMSE as the probe quantity was reduced for compressed fitting (Figure 9a,b), with larger increases occurring specifically below 12 probes for the Western data (Figure 9a). When performing singular value optimization tests using RMSE for initial inputs of 8, 16, and 32 probes, the RMSE was more sensitive to the addition of more singular values when fewer probes were used. When 32 probes were used, more singular values could be retained before seeing a noticeable increase in error (Figure 9c).

## 4 | Discussion

In this work, we introduced a field monitoring procedure aimed to accurately fit higher order phase accrual using fewer field probes than are typically required. The technique requires a calibration scan that involves rotations and/or translations of the field probe array around the imaging volume, followed by the creation of a compression matrix.

*Truncation trade-offs.* We found that the optimal number of singular vectors to retain in the compression matrix is generally less than the maximum permitted for any given number of probes. With too few singular values, RMSE and image degradation increases due to the removal of important basis functions. However, when the total number of basis functions is close to the number of probes, $\hat{P}$ becomes more poorly conditioned leading to increased error when solving Equation 9. As shown in this work, the "optimal" number of singular values can vary between scanners and with different numbers of probes, which depends on the degree of higher order information contained in the principal components. For example, the calibration data from ETH Zurich displayed better performance when 7 singular values were used, indicating that the high-performance gradient exhibits pronounced higher order behaviour,[21] and/or the conditioning of $\hat{P}$ is better suited to handle more basis functions (likely due to closer and more evenly distributed probe positions). While manual determination of the optimal number of singular values was determined by comparison to a ground truth with many probes, the development of less user-intensive means of determining the threshold could be an area of future work.

*Generalizability: resolution.* Despite possessing relatively different gradient amplitude and frequency profiles, minimal differences in k-coefficient fits and image quality were observed for resolution-varying calibration data. However, the best results were observed when using the high-resolution scan exhibiting a broader frequency content and a longer readout time. Thus, a high-bandwidth acquisition is recommended for calibration as it accesses a greater variety of higher spatial orders and improves generalizability. *Diffusion directions.* The consistently low RMSE values and comparable DWI when using 30 diffusion directions down to 6 for calibration suggests that good characterization can be performed if there are sufficient directions that are distributed uniformly to capture all eddy current self- and cross-terms (i.e., 6 degrees of freedom). *Diffusion gradient shape.* The observable differences in spatial distribution maps and RMSE values indicate that the eddy current modes excited by the PGSE and OGSE diffusion gradients are different. PGSE exhibits a broader spectral response compared to OGSE, which excites a finite frequency range. Additionally, OGSE waveforms result in reduced higher spatial eddy current modes given their partial canceling due to the alternating positive and negative gradient ramps.[10,43] As a result, OGSE calibration data is less generalizable than PGSE, which likely explains why the combined compressed basis function map strongly resembles the PGSE case, and why attempting to reconstruct PGSE data using the OGSE calibration leads to significant errors. *EPI vs spiral*. The poor performance of combined EPI/spiral calibration data for the spiral acquisition was surprising and suggests that there are limits in generalizability when the number of probes (and thus number of singular vectors) is limited. That is, the combined calibration likely requires retaining more compressed basis functions than we can fit with 16 probes. Accordingly, there is likely a limited ability to design a "one-size-fits-all" calibration scan that is appropriate for all acquisitions. *b-tensor encoding acquisition*. With the above findings in mind, a b-tensor acquisition was reconstructed using the compression matrix devised from "scan 1" (higher resolution calibration with PGSE diffusion gradients). Despite notable differences in imaging parameters between the acquisition and calibration data, the high-quality results show that the selected calibration data is appropriate to use for a variety of routine acquisitions. Specifically, the scan incorporated spherical tensor encoding diffusion gradients, but the high-image quality seen for STE images confirms that the PGSE calibration successfully captures the basis functions that are commonly expressed by the induced eddy currents. Moreover, this scan incorporated higher b-values than the ones included in the calibration, suggesting that accurate characterizations are still achievable with different eddy current scaling. In summary, though there are trade-offs with generalizability and accuracy (e.g., combined EPI/spiral calibration performed poorly), results showed that somewhat general calibration scans for "classes" of sequences (e.g., one compression matrix for all spiral diffusion MRI) may be possible. Nevertheless, the design of optimal calibration scans is an avenue for future work. This includes the exploration of calibration optimization when performing different scan geometries, including small axial tilts experienced along commonly imaged orientations such as the anterior commissure - posterior commissure line.

*Other sites*. The good agreement in k-coefficients and images relative to the ground truth cases achieved for two other scanners suggests that the technique is applicable to different scanners and gradient systems. The compression effectiveness of the Stanford calibration acquisition suggests that general acquisitions like chirped pulses have potential to identify many commonly excited basis functions, if all three gradient axes are probed. However, the application of this approach to routine imaging acquisition needs to be further explored given the generalizability/accuracy trade-offs described above.

*Fewer probes*. The investigation of fewer than 16 probes was motivated by the notion that less probes may be used to characterize the field dynamics, if the probe quantity is greater than or equal to the total number of basis function terms. As results from our system showed, fits remained stable until 12 probes were used, and for the ETH Zurich data, images maintained good quality down to 10 probes. Our data's higher sensitivity to probe reduction is likely due to the quicker ill-conditioning of the problem as the probe quantity approaches the number of basis functions, whereas ETH Zurich data is better conditioned due to probes being placed within the DSV. The potential to reduce the probe amount required for high order fitting below 16 provided commercially may be advantageous in certain scenarios. Namely, when integrating field probes into RF coils, challenges in probe placement can arise especially when handling complex coils that house many elements. Fewer probes may simplify the incorporation of all probes and required elements. Additionally, instances may arise where only select probes are subject to rapid field variations. This technique enables removal of unfavorable probes that may hinder phase fitting, while still permitting higher order characterizations.

*Comparison to other approaches*. The proposed technique was also compared to a previous approach presented by the authors, which in addition to weighted least squares fitting discussed earlier, fit the spatial orders one at a time to reduce error in lower order terms.[24] While application of this previous technique resulted in reduced RMSE error and DWI blurring, further significant improvements were observed using basis function compression (Figure S4).

*Limitations*. The calibration scheme did not consider spatial modes that might arise from physiological processes during concurrent monitoring, such as field changes from breathing. However, the differences are expected to be small due to the use of rapid single-shot trajectories. Additionally, for the Western data, a GIRF-like calibration scan was not considered. However, given that we already show challenges for generalizability here, it is unlikely that such a calibration scan would perform well for our system. It is also worth noting that acquiring field monitoring measurements in various probe positions may be challenging, particularly when using integrated field probe coils. For our configuration, modifications to cable connections and removal of the patient cradle were necessary to allow for relatively unrestricted movement of the head coil within the imaging volume. Lastly, it is likely that changing the RF coil position can alter the eddy current patterns due to reconfiguration of RF shielding components, particularly if asymmetric shielding is used. While not a large concern for the head-only scanner due to the absence of shielding in the head coil, this may be worth investigating on other systems by assessing the degree of field dynamic changes with reconfiguration of the RF coil.

## 5 | Conclusion

The presented field monitoring strategy enables the incorporation of additional higher order information than is permitted with a conventional collection of probes. This results in improvements in accuracy of field dynamics and diffusion data, specifically when field probes experience rapid spatial field variations. With the rise of specialized gradient systems having high gradient strengths and slew rates, field probes become more attractive to account for higher order perturbations. Using this technique to accurately measure these higher spatial orders may improve the ability to tap into the advanced capabilities of these systems, while maintaining high-quality image production.

### Acknowledgements

The authors would like to thank the NSERC Discovery Grant [grant number RGPIN-2018-05448], Canada Research Chairs [number 950-231993], Canada First Research Excellence Fund to BrainsCAN, and the NSERC CGS D program.

### Data Availability Statement

The basis function compression technique and the source code used for the expanded encoding model reconstructions are publicly available in the MatMRI toolbox: https://zenodo.org/records/5708265 and https://gitlab.com/cfmm/matlab/matmri. A demo calculating k-coefficients from raw probe data measured from a spiral acquisition on the head-only 7T with and without the proposed fitting scheme is included.

# References

1.  Barmet C, De Zanche N, Pruessmann KP. Spatiotemporal magnetic field monitoring for MR. *Magn Reson Med*. 2008;60(1):187-197.

2.  De Zanche N, Barmet C, Nordmeyer-Massner JA, Pruessmann KP. NMR probes for measuring magnetic fields and field dynamics in MR systems. *Magn Reson Med*. 2008;60(1):176-186.

3.  Barmet C, De Zanche N, Wilm BJ, Pruessmann KP. A transmit/receive system for magnetic field monitoring of in vivo MRI. *Magn Reson Med*. 2009;62(1):269-276.

4.  Sipilä P, Lange D, Lechner S, et al. Robust, susceptibility-matched NMR probes for compensation of magnetic field imperfections in magnetic resonance imaging (MRI). *Sens Actuators A Phys*. 2008;145-146:139-146.

5.  Vannesjo SJ, Wilm BJ, Duerst Y, et al. Retrospective correction of physiological field fluctuations in high-field brain MRI using concurrent field monitoring. *Magn Reson Med*. 2015;73(5):1833-1843.

6.  Boesch C, Gruetter R, Martin E. Temporal and spatial analysis of fields generated by eddy currents in superconducting magnets: optimization of corrections and quantitative characterization of magnet/gradient systems. *Magn Reson Med*. 1991;20(2):268-284.

7.  Stejskal EO, Tanner JE. Spin diffusion measurements: Spin echoes in the presence of a time-dependent field gradient. *J Chem Phys*. 1965;42(1):288-292.

8.  Schachter M, Does MD, Anderson AW, Gore JC. Measurements of restricted diffusion using an oscillating gradient spin-echo sequence. *J Magn Reson*. 2000;147(2):232-237.

9.  Baron CA, Kate M, Gioia LC, et al. Oscillating Gradient Spin-Echo (OGSE) DTI Yields Mechanistic Insights in Human Stroke. In: *Proceedings of the 22th Annual Meeting of ISMRM; Milano, Italy*. pdfs.semanticscholar.org; 2014:0106.

10. Valsamis JJ, Dubovan PI, Baron CA. Characterization and correction of time-varying eddy currents for diffusion MRI. *Magn Reson Med*. December 2021. doi:10.1002/mrm.29124

11. Wong EC, Cox RW, Song AW. Optimized isotropic diffusion weighting. *Magn Reson Med*. 1995;34(2):139-143.

12. Lampinen B, Szczepankiewicz F, Mårtensson J, van Westen D, Sundgren PC, Nilsson M. Neurite density imaging versus imaging of microscopic anisotropy in diffusion MRI: A model comparison using spherical tensor encoding. *Neuroimage*. 2017;147:517-531.

13. Le Bihan D, Poupon C, Amadon A, Lethimonnier F. Artifacts and pitfalls in diffusion MRI. *J Magn Reson Imaging*. 2006;24(3):478-488.

14. Jezzard P, Barnett AS, Pierpaoli C. Characterization of and correction for eddy current artifacts in echo planar diffusion imaging. *Magn Reson Med*. 1998;39(5):801-812.

15. Jensen JH, Helpern JA, Ramani A, Lu H, Kaczynski K. Diffusional kurtosis imaging: the quantification of non-gaussian water diffusion by means of magnetic resonance imaging. *Magn Reson Med*. 2005;53(6):1432-1440.

16. Lasič S, Szczepankiewicz F, Eriksson S, Nilsson M, Topgaard D. Microanisotropy imaging: quantification of microscopic diffusion anisotropy and orientational order parameter by diffusion MRI with magic-angle spinning of the q-vector. *Frontiers in Physics*. 2014;2. doi:10.3389/fphy.2014.00011

17. Szczepankiewicz F, Lasič S, van Westen D, et al. Quantification of microscopic diffusion anisotropy disentangles effects of orientation dispersion from microstructure: applications in healthy volunteers and in brain tumors. *Neuroimage*. 2015;104:241-252.

18. Mueller L, Wetscherek A, Kuder TA, Laun FB. Eddy current compensated double diffusion encoded (DDE) MRI. *Magn Reson Med*. 2017;77(1):328-335.

19. Wilm BJ, Barmet C, Pavan M, Pruessmann KP. Higher order reconstruction for MRI in the presence of spatiotemporal field perturbations. *Magn Reson Med*. 2011;65(6):1690-1701.

20. Wilm BJ, Barmet C, Gross S, et al. Single-shot spiral imaging enabled by an expanded encoding model: Demonstration in diffusion MRI. *Magn Reson Med*. 2017;77(1):83-91.

21. Wilm BJ, Hennel F, Roesler MB, Weiger M, Pruessmann KP. Minimizing the echo time in diffusion imaging using spiral readouts and a head gradient system. *Magn Reson Med*. 2020;84(6):3117-3127.

22. Huang SY, Witzel T, Keil B, et al. Connectome 2.0: Developing the next-generation ultra-high gradient strength human MRI scanner for bridging studies of the micro-, meso- and macro-connectome. *Neuroimage*. 2021;243:118530.

23. Feinberg DA, Beckett AJS, Vu AT, et al. Next-generation MRI scanner designed for ultra-high-resolution human brain imaging at 7 Tesla. *Nat Methods*. 2023;20(12):2048-2057.

24. Dubovan PI, Gilbert KM, Baron CA. A correction algorithm for improved magnetic field monitoring with distal field probes. *Magn Reson Med*. 2023;90(6):2242-2260.

25. Gilbert KM, Dubovan PI, Gati JS, Menon RS, Baron CA. Integration of an RF coil and commercial field camera for ultrahigh-field MRI. *Magn Reson Med*. December 2021. doi:10.1002/mrm.29130

26. Buehrer M, Pruessmann KP, Boesiger P, Kozerke S. Array compression for MRI with large coil arrays. *Magn Reson Med*. 2007;57(6):1131-1139.

27. Zhang T, Pauly JM, Vasanawala SS, Lustig M. Coil compression for accelerated imaging with Cartesian sampling. *Magn Reson Med*. 2013;69(2):571-582.

28. Addy NO, Wu HH, Nishimura DG. Simple method for MR gradient system characterization and k-space trajectory estimation. *Magn Reson Med*. 2012;68(1):120-129.

29. Vannesjo SJ, Haeberlin M, Kasper L, et al. Gradient system characterization by impulse response measurements with a dynamic field camera. *Magn Reson Med*. 2013;69(2):583-593.

30. Jones DK, Horsfield MA, Simmons A. Optimal strategies for measuring diffusion in anisotropic systems by magnetic resonance imaging. *Magn Reson Med*. 1999;42(3):515-525.

31. Arbabi A, Kai J, Khan AR, Baron CA. Diffusion dispersion imaging: Mapping oscillating gradient spin-echo frequency dependence in the human brain. *Magn Reson Med*. 2019;(July):1-12.

32. Wilm BJ, Nagy Z, Barmet C, et al. Diffusion MRI with concurrent magnetic field monitoring. *Magn Reson Med*. 2015;74(4):925-933.

33. Varela-Mattatall G, Dubovan PI, Santini T, Gilbert KM, Menon RS, Baron CA. Single-shot spiral diffusion-weighted imaging at 7T using expanded encoding with compressed sensing. *Magn Reson Med*. 2023;90(2):615-623.

34. Huang F, Vijayakumar S, Li Y, Hertel S, Duensing GR. A software channel compression technique for faster reconstruction with many channels. *Magn Reson Imaging*. 2008;26(1):133-141.

35. King SB, Varosi SM, Duensing GR. Optimum SNR data compression in hardware using an Eigencoil array. *Magn Reson Med*. 2010;63(5):1346-1356.

36. Uecker M, Lai P, Murphy MJ, et al. ESPIRiT--an eigenvalue approach to autocalibrating parallel MRI: where SENSE meets GRAPPA. *Magn Reson Med*. 2014;71(3):990-1001.

37. Dubovan PI, Baron CA. Model-based determination of the synchronization delay between MRI and trajectory data. *Magn Reson Med*. 2023;89(2):721-728.

38. Larsson EG, Erdogmus D, Yan R, Principe JC, Fitzsimmons JR. SNR-optimality of sum-of-squares reconstruction for phased-array magnetic resonance imaging. *J Magn Reson*. 2003;163(1):121-123.

39. Tournier JD, Smith R, Raffelt D, et al. MRtrix3: A fast, flexible and open software framework for medical image processing and visualisation. *Neuroimage*. 2019;202:116137.

40. Veraart J, Novikov DS, Christiaens D, Ades-Aron B, Sijbers J, Fieremans E. Denoising of diffusion MRI using random matrix theory. *Neuroimage*. 2016;142:394-406.

41. Arezza NJJ, Tse DHY, Baron CA. Rapid microscopic fractional anisotropy imaging via an optimized linear regression formulation. *Magn Reson Imaging*. 2021;80:132-143.

42. Weiger M, Overweg J, Rösler MB, et al. A high-performance gradient insert for rapid and short-T2 imaging at full duty cycle. *Magn Reson Med*. 2018;79(6):3256-3266.

43. Chan RW, Von Deuster C, Giese D, et al. Characterization and correction of eddy-current artifacts in unipolar and bipolar diffusion sequences using magnetic field monitoring. *J Magn Reson*. 2014;244:74-84.

| Scan Description | Resolution | TE | TR | Bandwidth | Diffusion Weighting |
|---|---|---|---|---|---|
| *Scan 1*<br>Higher res spiral PGSE | 1.3 x 1.3 x 2 mm$^2$ | 39 ms | 2,500 ms | 2014 Hz/Px | PGSE encoding<br>b = 1000 s/mm$^2$ |
| *Scan 2*<br>Lower res spiral PGSE | 2 x 2 x 2 mm$^2$ | 39 ms | 2,500 ms | 4006 Hz/Px | PGSE encoding<br>b = 1000 s/mm$^2$ |
| *Scan 3*<br>Spiral OGSE | 1.3 x 1.3 x 2 mm$^2$ | 93 ms | 2,500 ms | 2014 Hz/Px | OGSE encoding<br>b = 400 s/mm$^2$<br>frequency = 40 Hz |
| *Scan 4*<br>EPI PGSE | 1.3 x 1.3 x 2 mm$^2$ | 53 ms | 2,500 ms | 2140 Hz/Px | PGSE encoding<br>b = 1000 s/mm$^2$ |

**Table 1** Calibration Acquisition Details. To assess the impact of readout gradient frequency content on basis function compression, identical single-shot spiral diffusion-weighted acquisitions with different in-plane resolutions and resulting bandwidths were performed. Additionally, an OGSE scheme using 40 Hz oscillating diffusion gradients[10,31] was conducted to explore how frequency content of diffusion gradients may affect compression. Lastly, an EPI scan with similar imaging parameters as the higher resolution spiral acquisition was performed to explore the compression performance when calibrating with a different trajectory.

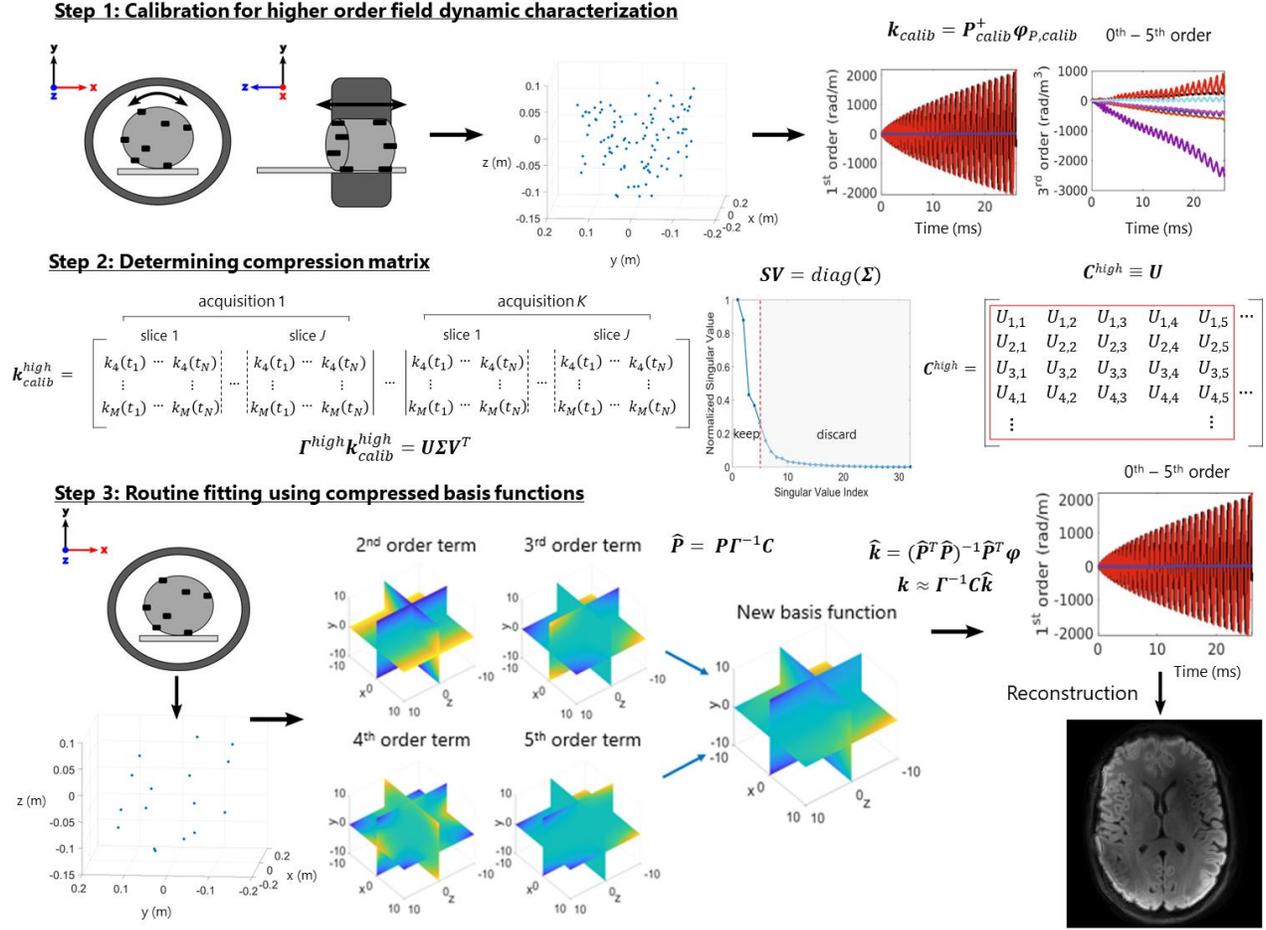

**Figure 1** An outline of the typical compression process. In the first step, higher order field dynamics are characterized using a compilation of probe measurements acquired using various probe positions. Principal component analysis is then performed on the desired higher order phase coefficient time-courses, from which a compression matrix is determined in step 2. The matrix is truncated based on inspection to preserve only the most relevant basis functions. Lastly, the compression matrix can be applied to compress basis functions sampled using the conventional probe arrangement, from which decompressed higher order field dynamics can be retrieved for use in an expanded encoding model-based reconstruction.

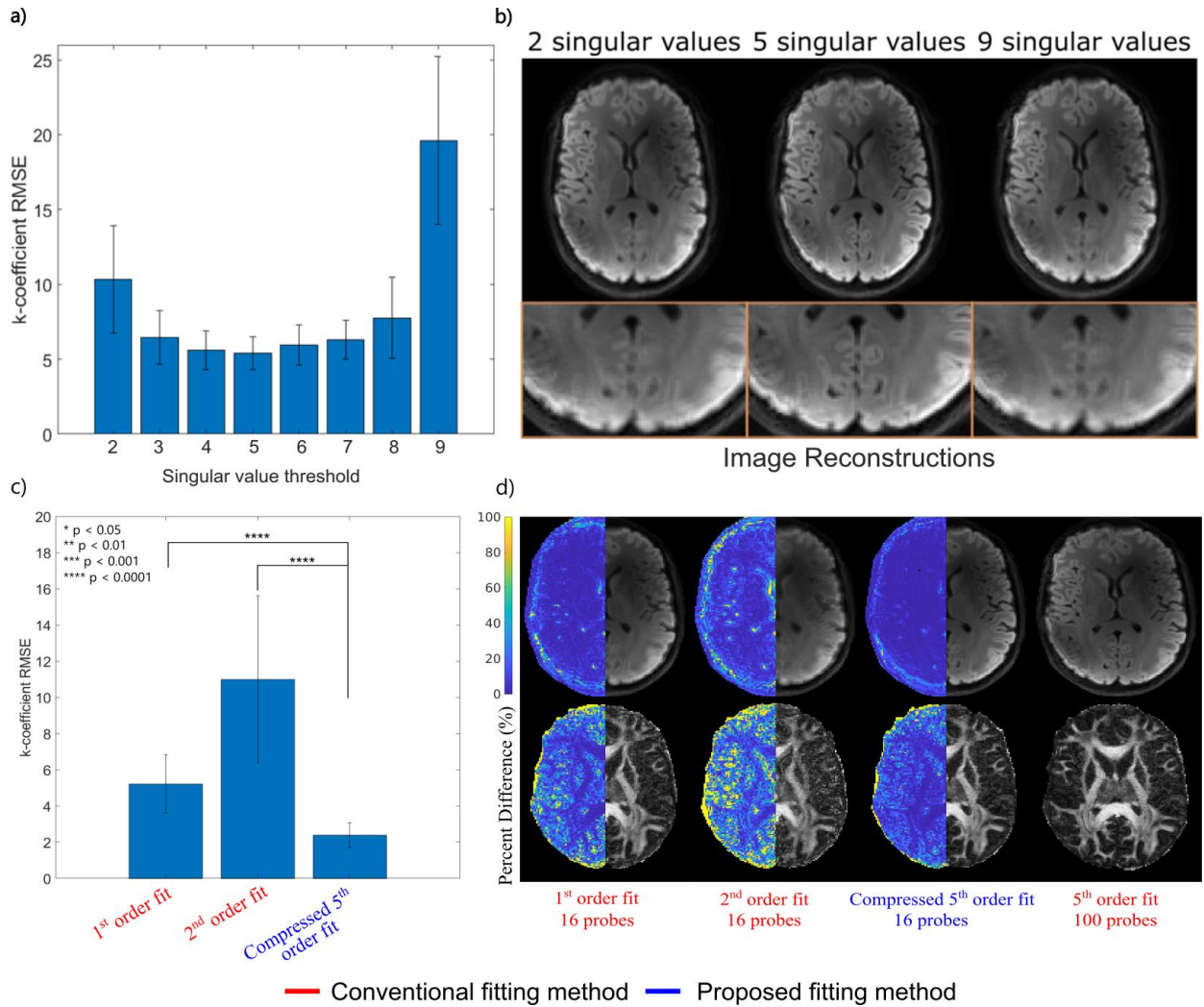

**Figure 2** Investigation of singular value threshold compression performance for scan 1. Shown in (a) is the average RMSE of up-to-second-order computed k-coefficients relative to the ground truth as a function of the singular value quantity preserved for compression. (b) presents the reconstructed mean DWI informed by the compressed k-coefficients using the outlined number of singular values. Zoom-ins highlight the substantial blurring experienced at the low and high singular value regimes. c) illustrates the error for up-to-first-order k-coefficients relative to the ground truth (fifth order fit using 100 probes) for conventional first and second order fits and compressed fifth order fit using 5 singular values for compression. d) Respective reconstructed mean DWI and calculated FA maps when incorporating the same fitting schemes. Percent difference images were calculated relative to images informed by fifth order field dynamics and are shown in the left hemisphere of the images. Comparisons with conventional third order fits are not shown due to the absence of an overdetermined state and evidently higher errors exhibited (Video S1).

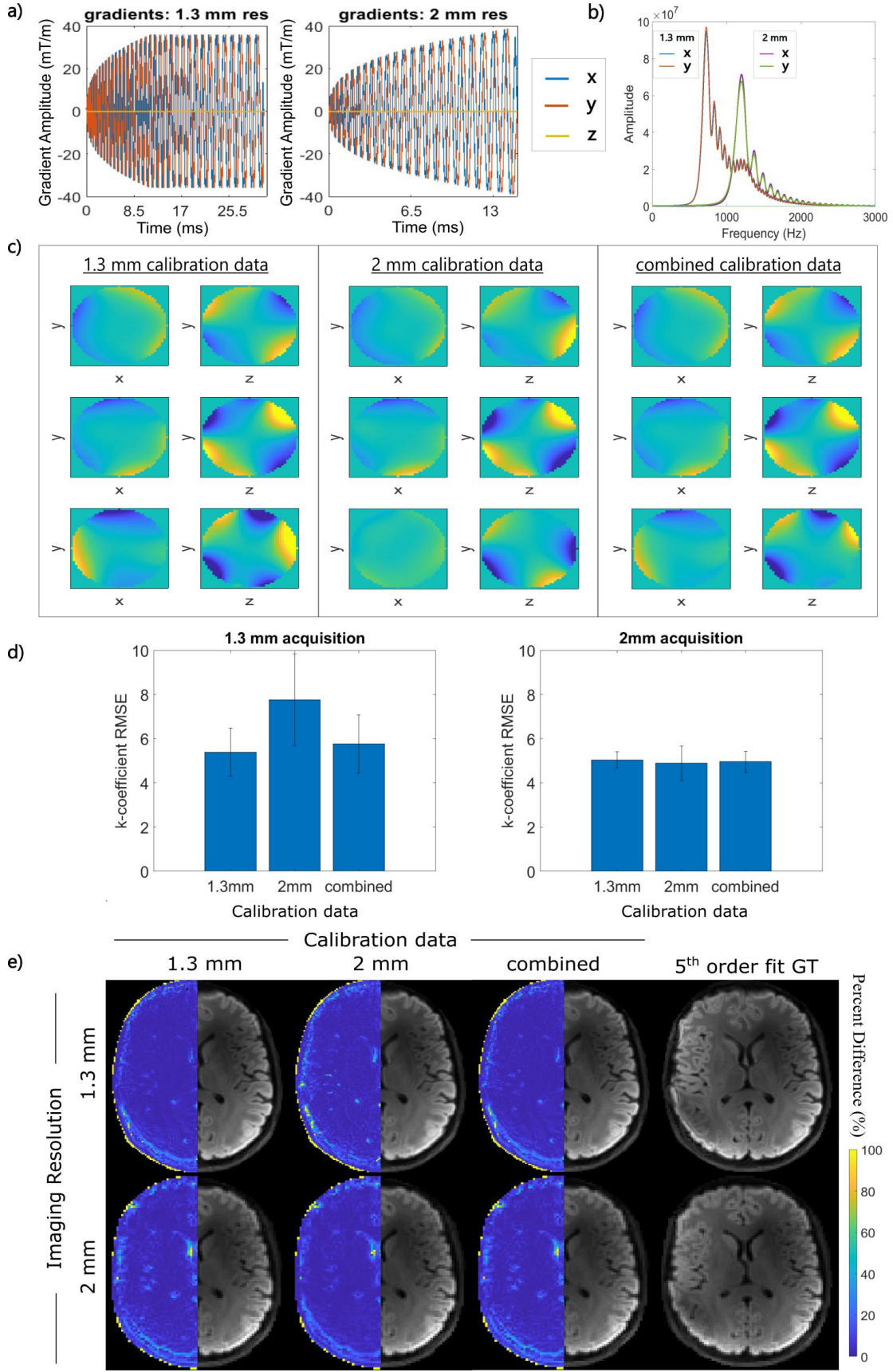

**Figure 3** (a) Gradient readout profiles for the 1.3-mm and 2-mm in-plane acquisitions, and (b) respective power spectral density profiles. (c) Cross-sectional spatial distribution (left to right: x-y, y-z planes) of compressed basis functions when using low, high, and combined resolutions for the calibration data. Rows represent the compressed basis functions related to the first three principal components. (d) Mean RMSE analysis of up-to-second-order k-coefficients, using low, high, and combined resolutions for the calibration data. Comparison was performed for field dynamics measured for the 1.3-mm acquisition (left) and 2-mm acquisition (right). (e) Mean DWI reconstructions for the 1.3-mm (top) and 2-mm (bottom) acquisitions, when informed by field dynamics compressed based on the described calibration data: left-to-right 1.3-mm acquisition, 2-mm acquisition, combined resolutions, plus respective fifth order fits calculated using 100 probes. Percent difference images were calculated relative to images informed by fifth order field dynamics and are shown in the left hemisphere of the images. 5 singular values were kept for compression.

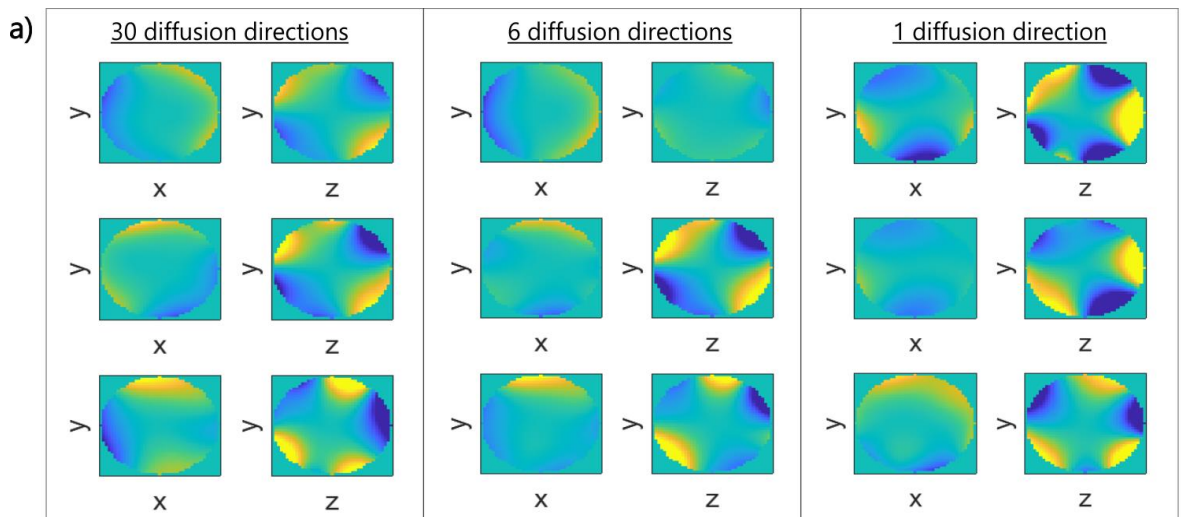

a)

30 diffusion directions    6 diffusion directions    1 diffusion direction

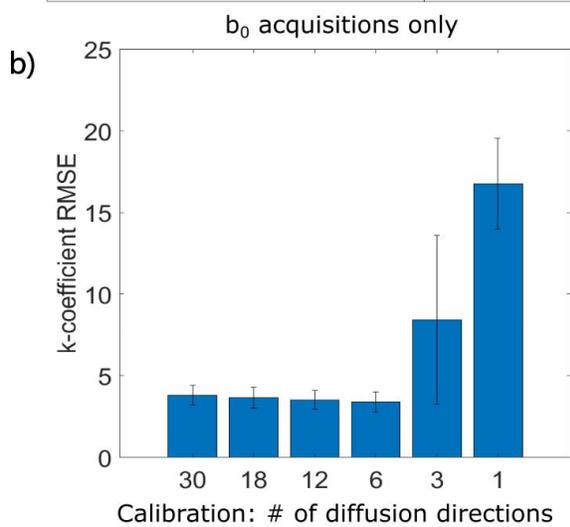

b) b₀ acquisitions only

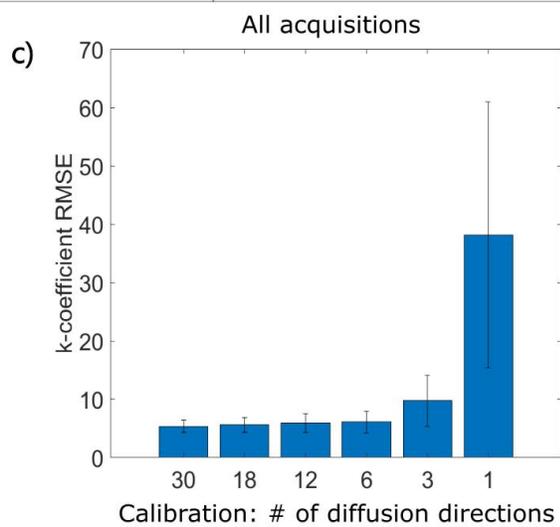

c) All acquisitions

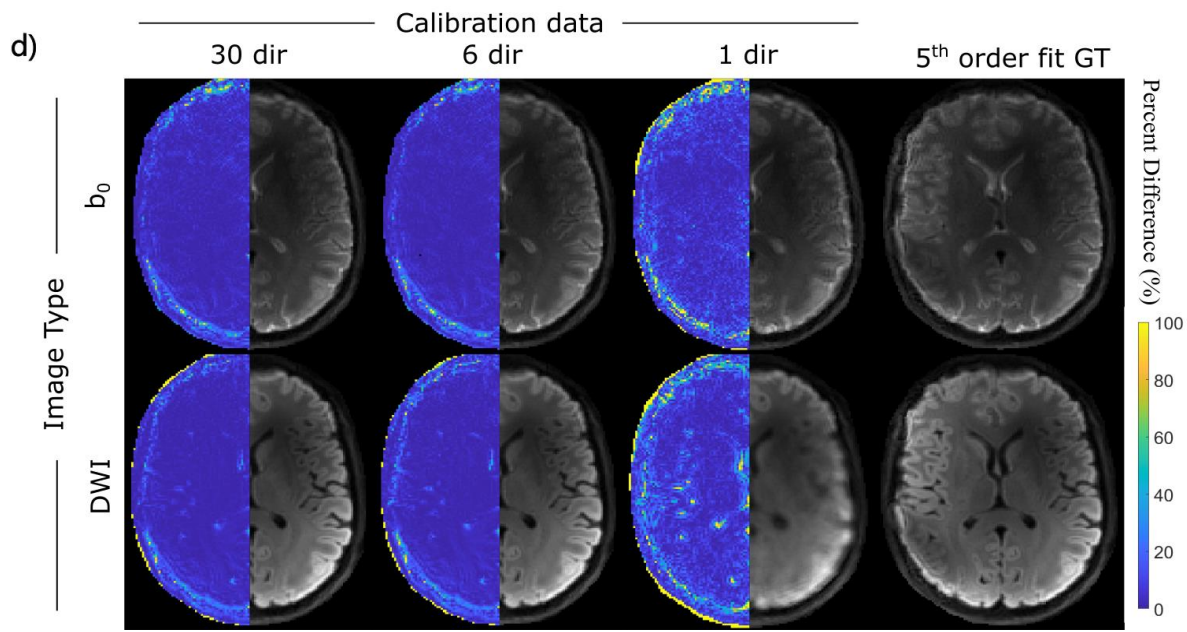

d)

**Figure 4** (a) Cross-sectional spatial distribution (left to right: x-y, y-z planes) of compressed basis functions for calibration data equipped only with diffusion data containing 30,6,1 directions. $b_0$ acquisitions were omitted from the calibration data itself for all relevant analysis. Rows represent the new basis functions related to the first three principal components. Mean RMSE analysis of up-to-second-order k-coefficients, when preserving the following amount of uniform diffusion directions in calibration: 30,18,12,6,3,1, for (b) $b_0$ acquisitions only, and (c) when including the diffusion acquisitions in the analysis. (d) Mean $b_0$ (top) and DWI (bottom) reconstructions when informed by field dynamics compressed using (left-to right) 30,6,1 diffusion directions in the calibration data, plus respective fifth order fits calculated using 100 probes. Percent difference images were calculated relative to images informed by fifth order field dynamics and are shown in the left hemisphere of the images. 5 singular values were kept for compression. Directions in the subsets were chosen to maximize electrostatic repulsion for each subset.

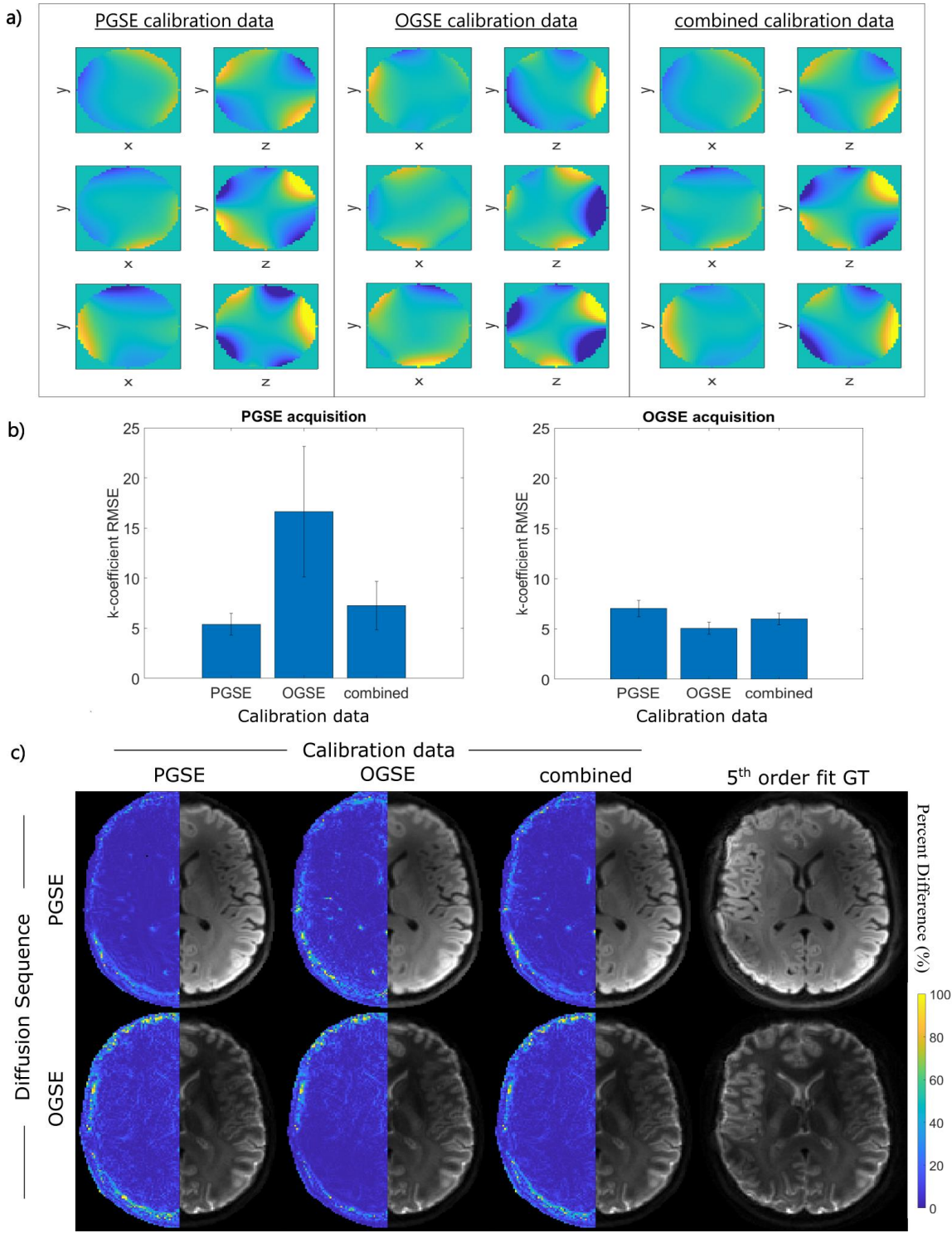

**Figure 5** (a) Cross-sectional spatial distribution (left to right: x-y, y-z planes) of compressed basis functions acquired from calibration data including PGSE, OGSE, and combined acquisitions. Rows represent the new basis functions related to the first three principal components. (b) Mean RMSE analysis of up-to-second-order k-coefficients, calculated using the calibration data defined from PGSE, OGSE, and combined acquisitions. Comparison was performed for field dynamics measured for the PGSE acquisition (left) and OGSE acquisition (right). (c) Mean DWI reconstructions for the PGSE (top) and OGSE (bottom) acquisitions, when informed by field dynamics compressed based on the described calibration data: left-to-right PGSE acquisition, OGSE acquisition, and combined diffusion data, plus respective fifth order fits calculated using 100 probes. Percent difference images were calculated relative to images informed by fifth order field dynamics and are shown in the left hemisphere of the images. 5 singular values were kept for compressions, except for calibrations containing only OGSE data, where 3 singular values were preserved.

a)

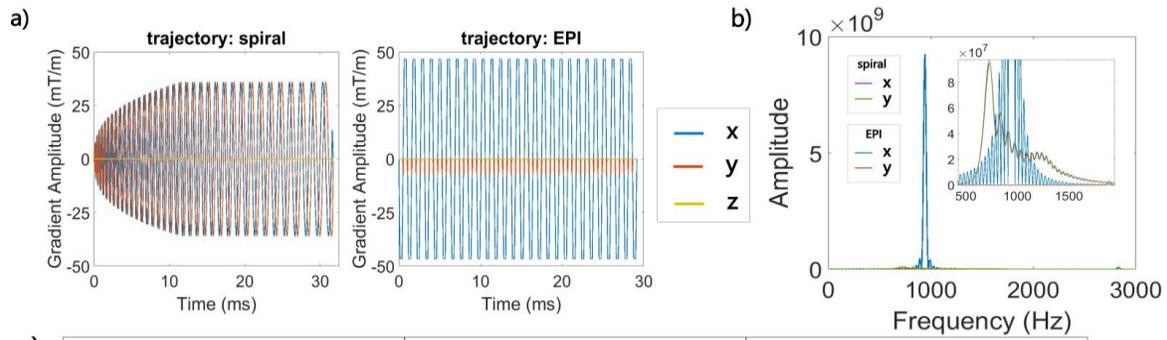

b)

c)

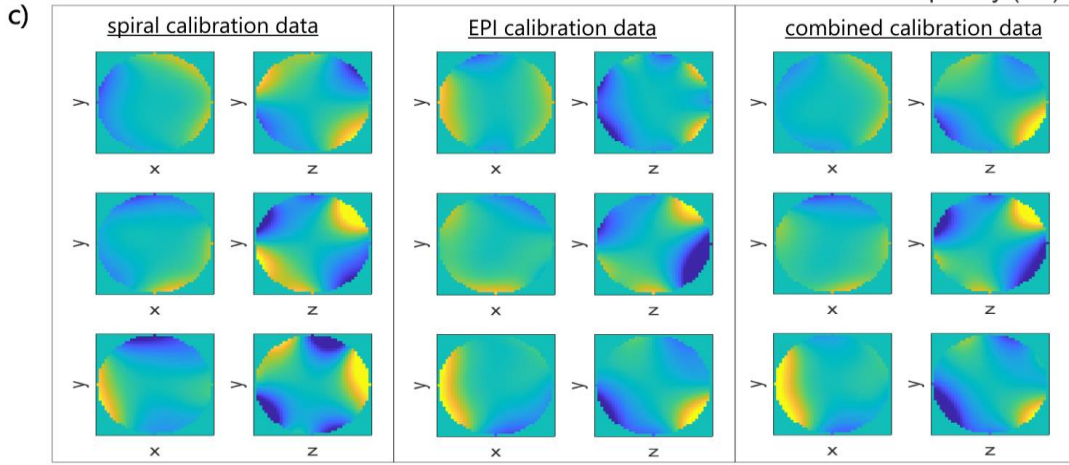

spiral calibration data     EPI calibration data     combined calibration data

d)

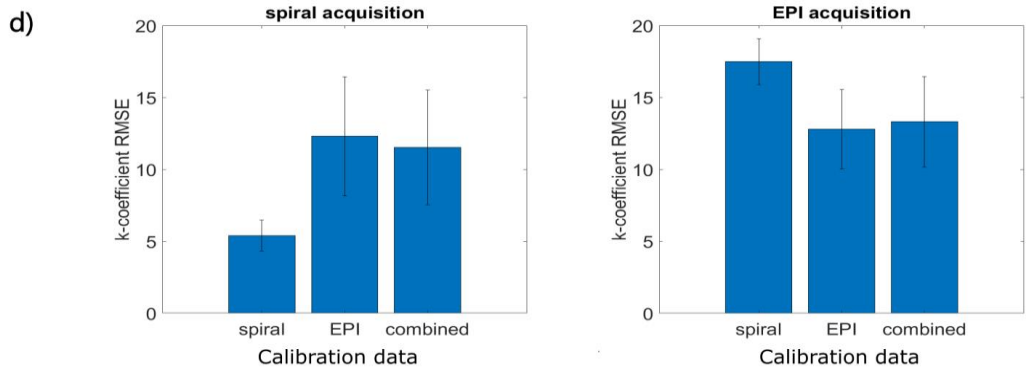

e)

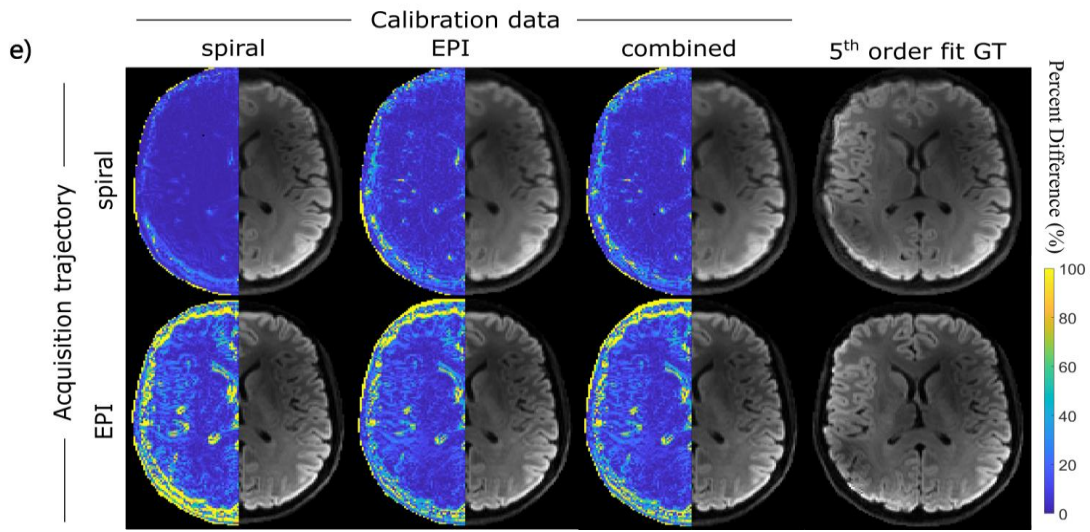

**Figure 6** (a) Gradient readout profiles for the spiral and EPI acquisitions, and (b) respective power spectral density profiles. (a) Cross-sectional spatial distribution (left to right: x-y, y-z planes) of compressed basis functions determined from calibration data using the spiral trajectory, EPI trajectory, and combined trajectories. Rows represent the new basis functions related to the first three principal components. (b) Mean RMSE analysis of up-to-second-order k-coefficients, calculated using the calibration data from spiral, EPI, and combined acquisitions. Comparison was performed for field dynamics measured for the spiral acquisition (left) and EPI acquisition (right). (c) Mean DWI reconstructions for the spiral (top) and EPI (bottom) acquisitions, when informed by field dynamics compressed based on the described calibration data: left-to-right spiral acquisition, EPI acquisition, combined trajectories, plus respective fifth order fits calculated using 100 probes. Percent difference images were calculated relative to images informed by fifth order field dynamics and are shown in the left hemisphere of the images. 5 singular values were kept for compressions, except for spiral reconstructions calibrated using EPI data, which preserved 4 singular values.

a)

| b = 0 | LTE b=150 | LTE b=1000 | LTE b=2000 | STE b=2000 | ADC | μFA |

b)

LTE b=2000

2nd order fit

Compressed 5th order fit

— Conventional fitting method    — Proposed fitting method

**Figure 7** (a) Mean sample image slices reconstructed for the linear and spherical tensor-encoded acquisitions incorporated in the b-tensor encoding scan, as well as computed ADC and µFA maps. Reconstructions were informed with compressed fifth order field dynamics, using the compression matrix determined from "Scan 1" calibration data. (b) Comparison of mean DWI from the LTE scan (b = 2000 s/mm$^2$) for reconstructions informed with conventional second order fits (top) and compressed fifth order fits (bottom), with zoom-ins highlighting the blurring reduction observed when implementing compressed 5$^{th}$ order fits.

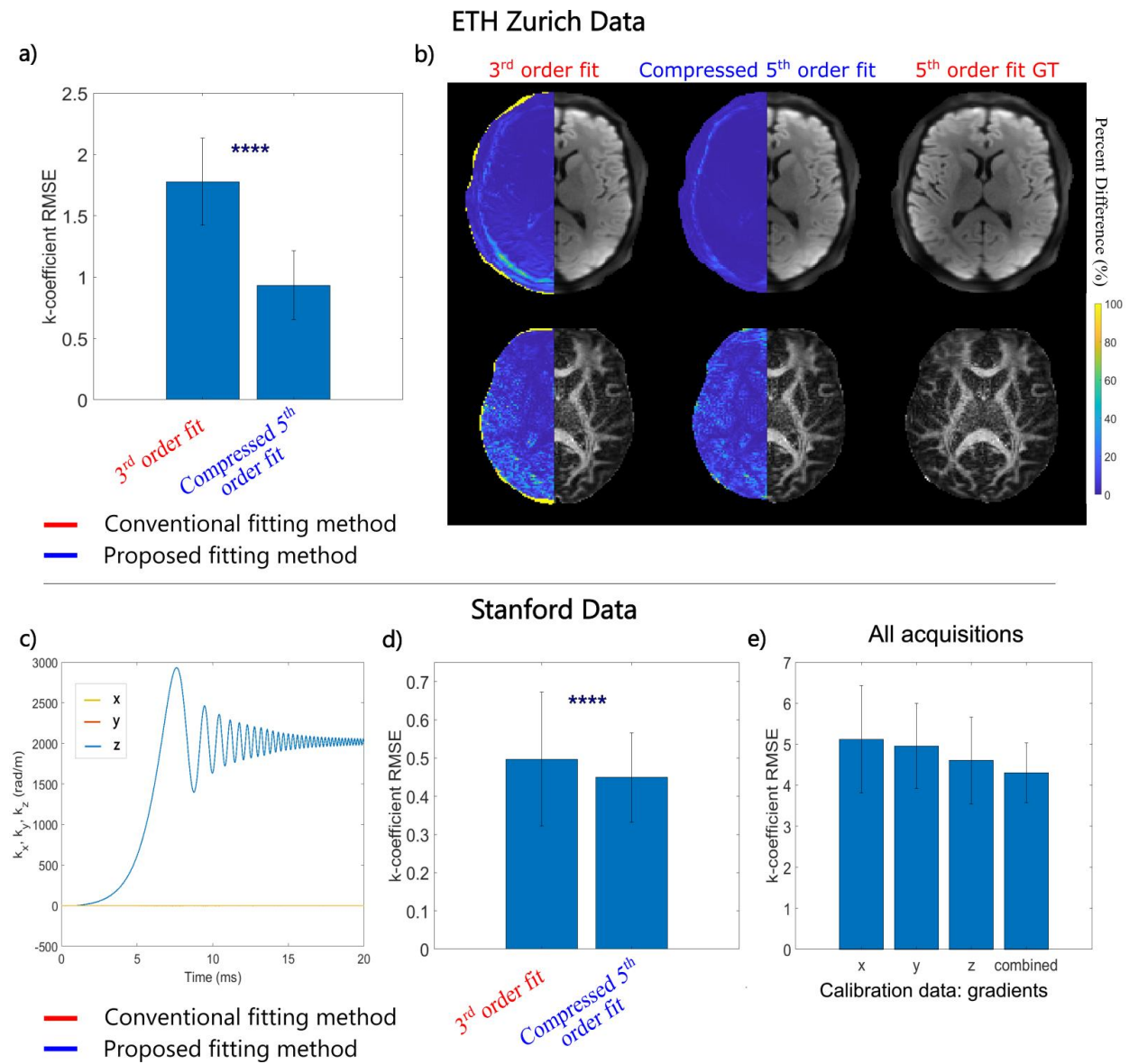

**Figure 8** Algorithm performance on data from two additional scanners: Philips 3T housing a high-performance gradient coil (top), and GE 3T Ultra-High-Performance scanner (bottom). (a) Mean RMSE comparison of k-coefficients up-to-second order when performing a conventional third order fit, and a compressed fifth order fit, relative to fifth order field dynamics computed using 64 probes. (b) Reconstructed mean DWI and calculated FA maps when informed by field dynamics determined by the same fitting schemes. Percent difference images were calculated relative to images informed by fifth order field dynamics and are shown in the left hemisphere of the images. 7 singular values were kept for compression. (c) First-order field-monitored trajectories from the chirped acquisition. (d) Mean RMSE comparison of k-coefficients up-to-second order when performing a conventional third order fit, and a compressed fifth order fit, relative to fifth order field dynamics computed using 48 probes. (e) Mean RMSE of k-coefficients as a function of the gradient frequency sweep acquisitions (x,y,z) included in the calibration data. 4 singular values were kept for compression.

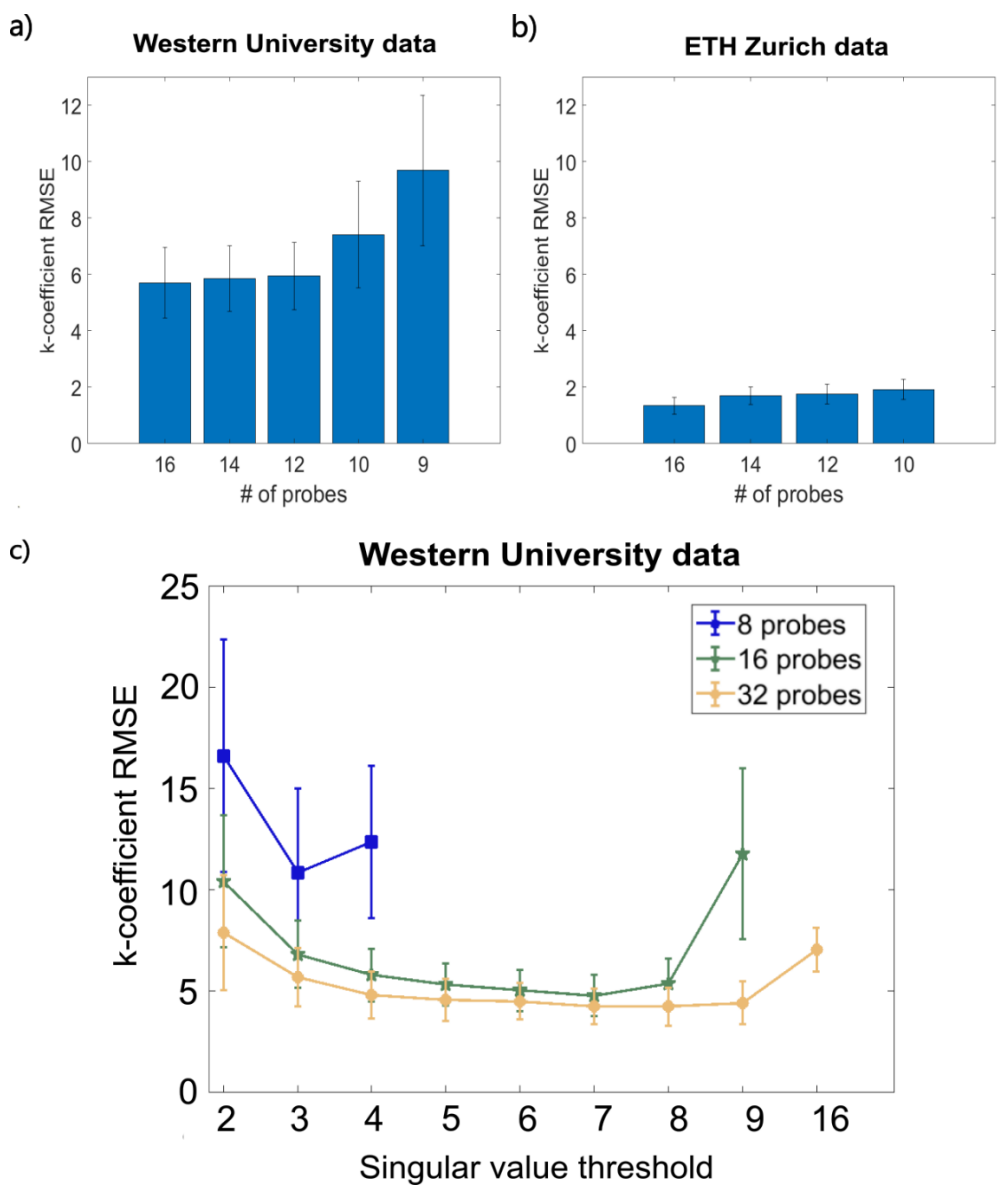

**Figure 9** Compression evaluation based on probe input for Western University and ETH Zurich data. (a) k-coefficient mean RMSE analysis for a range of 9-16 probes used for compressed fitting for Western University data. 5 singular values kept resulting in a total of 9 basis functions, hence the lowest probe amount equaling 9. (b) k-coefficient mean RMSE analysis for a range of 10-16 probes used for compressed fitting for ETH Zurich data. 6 singular values kept resulting in a total of 10 basis functions, hence the lowest probe amount equaling 10. (c) k-coefficient RMSE evaluation as a function of singular value quantity, for probe fitting amounts of 8,16, and 32 probes. Further points are not plotted for 8 probes and 16 probes due to the total number of basis functions exceeding the number of probes. Probe subsets were determined by maximizing electrostatic repulsion for each subset.

**Supporting Information**

**Supporting Information S1** Scan details for additional sites

ETH Zurich Data

A healthy volunteer was scanned on a 3T Philips Achieva scanner housing a high-performance head-only gradient insert (200 mT/m max gradient strength and 600 T/m/s max slew rate, [22, 22, 20 cm] imaging volume in x, y, z directions respectively)[42] after approval from the institutional review board. A single-shot spiral diffusion-weighted acquisition was performed with the following imaging parameters: FOV: 220 x 220 $mm^2$, 1.5 mm in-plane resolution, 3 mm slice thickness, number of slices: 20, TE/TR: 21/6,000 ms, rate 2 undersampling, bandwidth = 222 kHz, b = 0 $s/mm^2$ acquisitions: 2, diffusion directions: 6, b-value: 1000 $s/mm^2$ using a PGSE scheme, axial orientation. Data for coil sensitivities and $B_0$ maps was acquired using a multi-echo gradient echo scan. Using a dynamic field camera (Skope, Zurich, Switzerland), field monitoring of an identical imaging acquisition was performed in 4 different field camera positions, with each position involving a rotation of the camera about the z-axis, and one position being slightly translated along the z-direction. From this, a probe array consisting of 64 probes was compiled. A ground truth fifth-order fit was performed for the acquisition, followed by the determination of a truncated compression matrix. Using this matrix, a compressed fifth order fit and a conventional third order fit were performed using only the 16 probes from the first orientation, mean probe distance from isocenter = 8.4 cm (range: 7.7–8.9 cm). Images were reconstructed using an in-house model-based reconstruction algorithm at ETH Zurich.[1]

Stanford Data

A chirped scan was performed on the GE 3T Ultra-High-Performance scanner (100 mT/m max gradient strength, 200 T/m/s max slew rate, 50-cm imaging volume) at Stanford University's Center for Cognitive and Neurobiological Imaging. The scan involved repeated acquisitions of frequency modulations from 0 to 26 kHz individually along the x, y, and z gradients. A total of 48 sweeps were repeated along each gradient axis. Using a dynamic field camera (Skope, Zurich, Switzerland), field monitoring of the acquisition was performed in an empty scanner repeatedly for three different probe orientations, where the field probes were rotated about the z-axis in each case. Using this data, a probe array consisting of 48 probes was compiled, and a ground truth fifth order fit of the acquisition was performed. A compression matrix was determined using the calibration data. The singular value threshold was determined based on minimized RMSE of the k-coefficients relative to the ground truth. While frequency sweeps were 75 ms in duration, only data points up to 20 ms were included in the calibration as this comprised much of the high-amplitude gradient oscillations, and limited the amount of noisy data points that could propagate into the principal component analysis. A compressed fifth order fit and a conventional third order fit were performed using the conventional 16-probe arrangement, mean probe distance from isocenter = 8.5 cm (range: 7.9–8.8 cm).

**References**

1. Wilm BJ, Barmet C, Pavan M, Pruessmann KP. Higher order reconstruction for MRI in the presence of spatiotemporal field perturbations. *Magn Reson Med*. 2011;65(6):1690-1701.

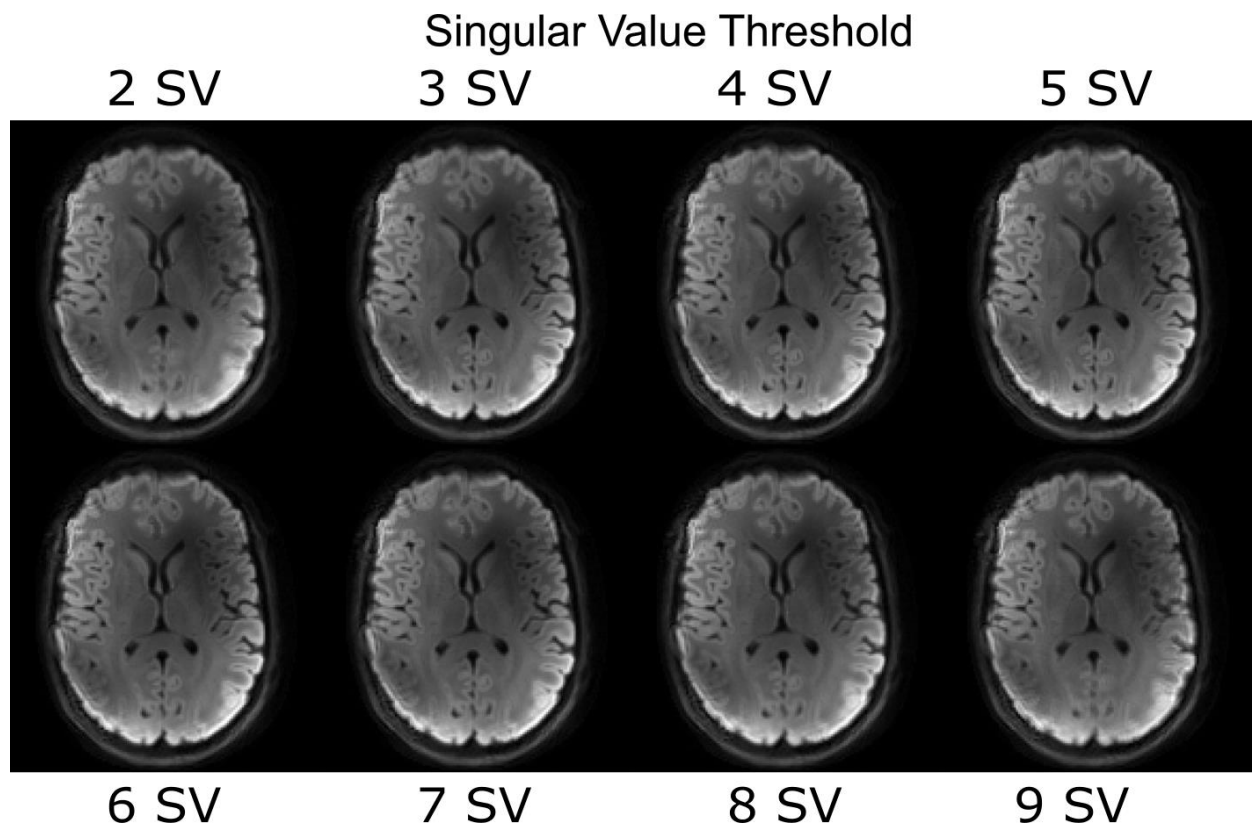

**Figure S1** Reconstructed mean DWI informed by compressed k-coefficients calculated from "Scan 1" calibration data, using the described number of singular values, for the complete range of 2-9 singular values investigated. Comparable image quality was observed in the range of 4-7 singular values, whereas use of 2-3 and 8-9 singular values introduced significantly more blurring.

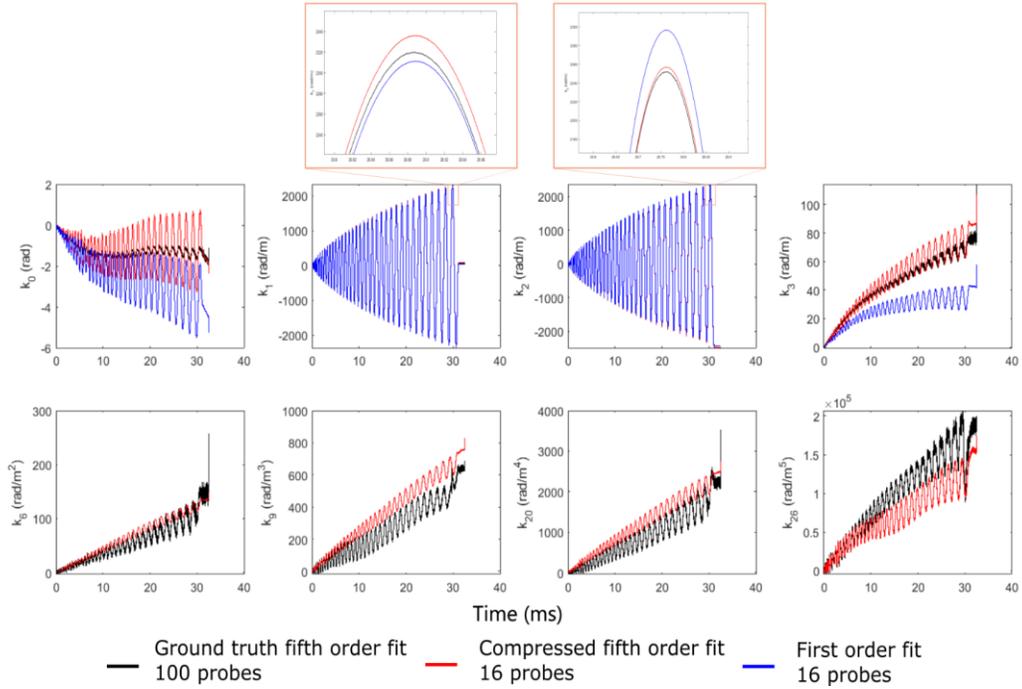

**Figure S2** Sample 0<sup>th</sup>-5<sup>th</sup> order k-coefficient time-courses of the 1.3-mm single-shot spiral acquisition (Scan 1), for different fitting methods: fifth order fit using 100 field probes (black), compressed fifth order fit (red), and conventional first order fit (blue). Overall, better agreement in first order terms was observed between the ground truth fifth order fit and compressed fifth order fit methods. Good agreement in these methods was also seen for the higher order terms.

## ETH Zurich Data

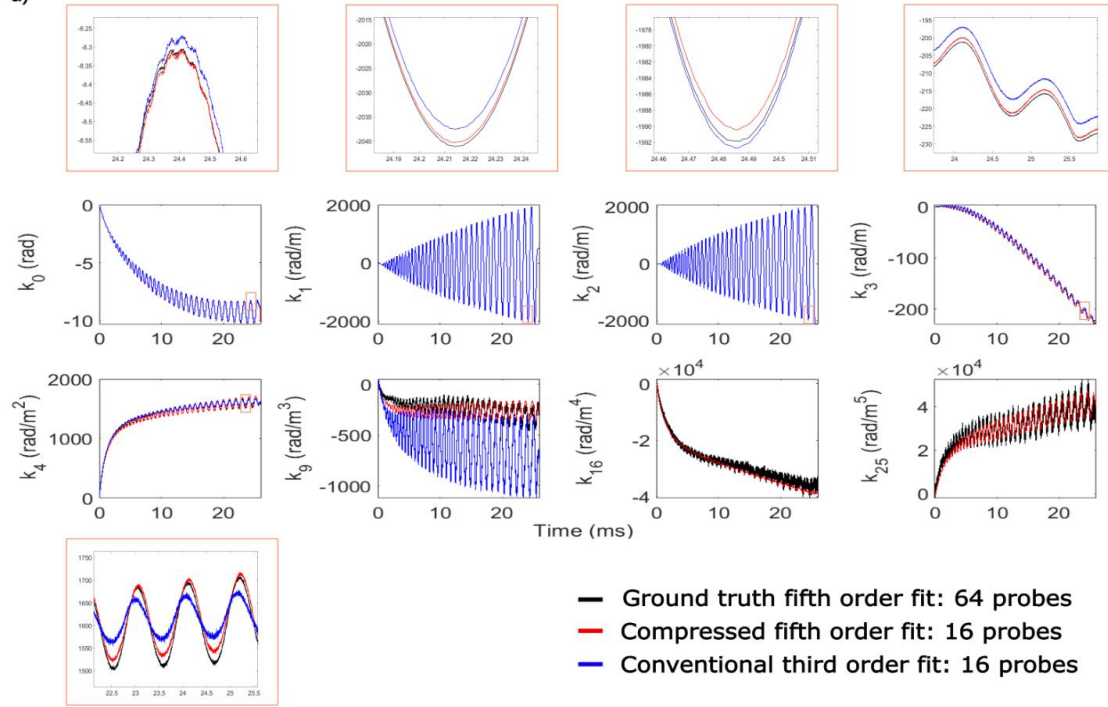

Ground truth fifth order fit: 64 probes
Compressed fifth order fit: 16 probes
Conventional third order fit: 16 probes

## Stanford Data

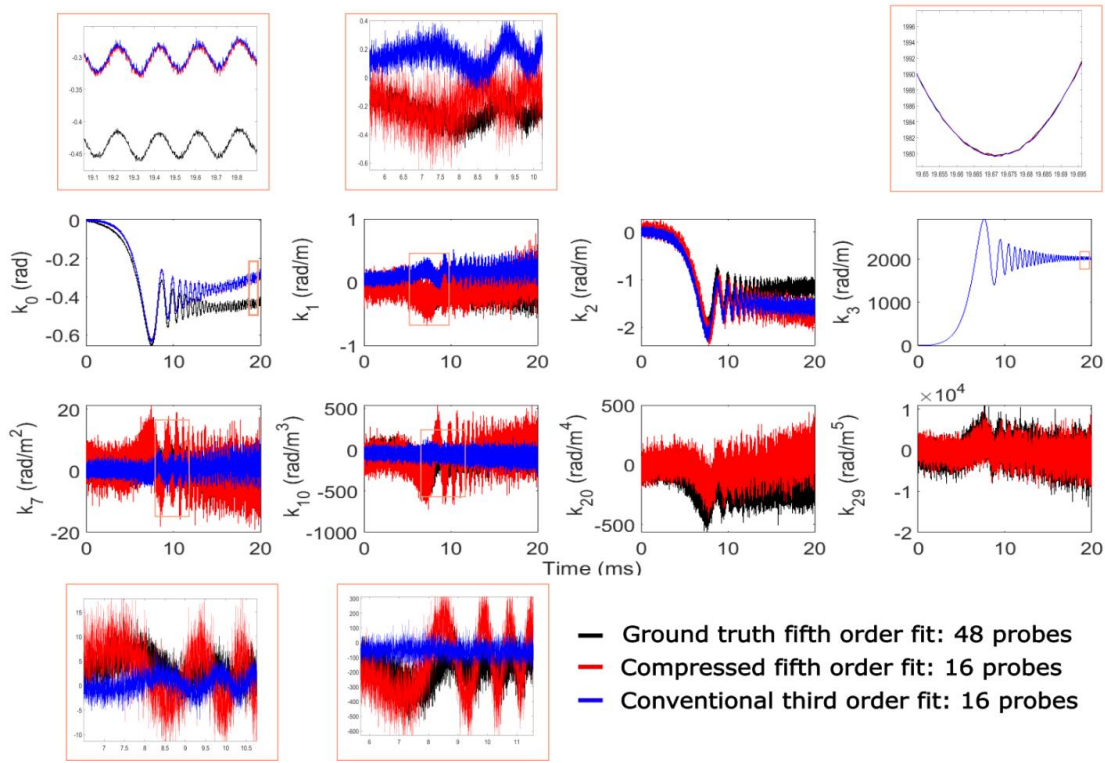

Ground truth fifth order fit: 48 probes
Compressed fifth order fit: 16 probes
Conventional third order fit: 16 probes

**Figure S3** Sample $0^{th}$-$5^{th}$ order k-coefficient time-courses of the single-shot spiral acquisition from (a) ETH Zurich and (b) the frequency sweep (Stanford), for different calculation methods: fifth order fit using 64 or 48 field probes (black), compressed fifth order fit (red), and conventional first order fit (blue). In both cases, better overall agreement was observed between the ground truth and compressed fifth order fit techniques, especially for second and third order terms.

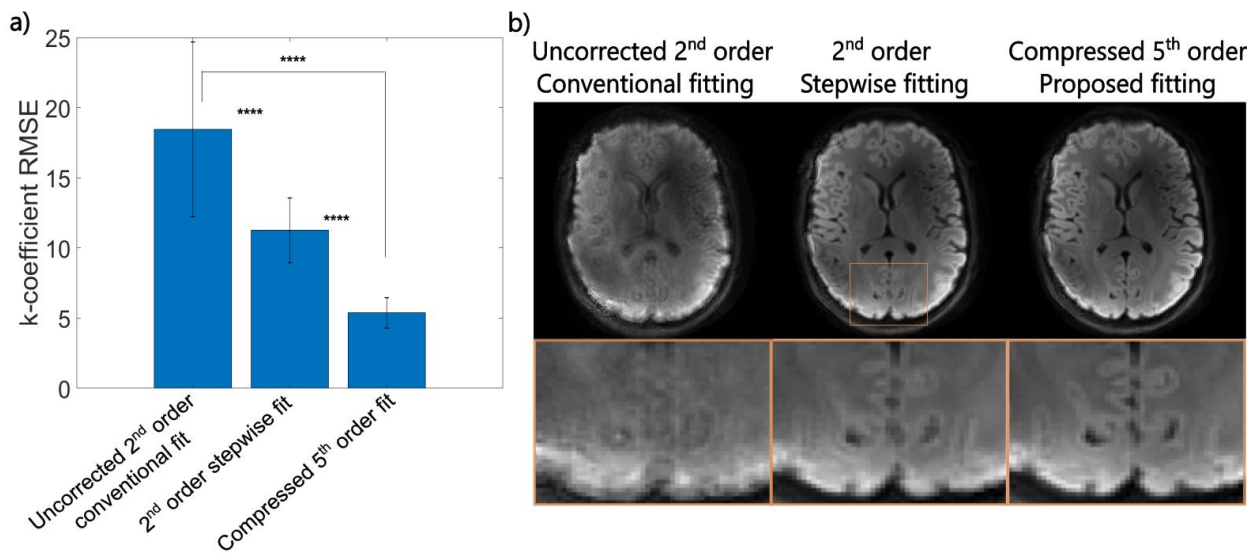

**Figure S4** Comparison of different proposed fitting techniques. (a) Quantitative k-coefficient RMSE analysis up-to-second order and (b) resulting qualitative mean DWI comparison informed by field dynamics computed conventionally with no form of fitting correction (left), using a previous fitting approach proposed by the authors (middle), and using the compressed basis function fitting approach (right). Improvements in both k-coefficient similarity and blurring reduction were observed with each successive iteration of fitting algorithm implemented.

**Video S1** Video comparison of mean DWI and FA maps informed by conventional third order fits using 16 probes, third order fit using 100 probes, and fifth order fit using 100 probes. Significant improvements in image quality were apparent when moving to third order fit with a probe surplus. Small reductions in blurring in the DWI and improvement in the FA map were also observed when implementing a fifth order fit.

**Video S2** Mean DWI from a spiral PGSE acquisition informed with calibration data from the PGSE acquisition, OGSE acquisition, and when combining the two diffusion acquisitions. Blurring becomes more apparent when the compression matrix was determined using data from the OGSE acquisition.